\documentclass[12pt]{article}

\usepackage{amsmath}
\usepackage{graphicx,psfrag,epsf}
\usepackage{enumerate}
\usepackage{natbib}
\usepackage{url} 

\usepackage{amsmath}

\usepackage{times}
\usepackage{bm}
\usepackage{natbib}
\usepackage{multirow}

\usepackage{amsmath,amssymb}
\usepackage{caption}
\usepackage{kotex}
\usepackage{color}

\usepackage[plain,noend]{algorithm2e}

\newcommand{\blind}{1}

\DeclareGraphicsExtensions{.jpg,.pdf,.png,.gif}
\renewcommand{\baselinestretch}{1.5}
\usepackage{amssymb}
\newtheorem{remark}{Remark}
\usepackage{multirow}

\newcommand{\bh}{\mathbf{h}}
\newcommand{\bx}{\mbox{\boldmath{$x$}}}

\newcommand{\bz}{\mathbf{z}}

\newcommand{\bbeta}{\boldsymbol{\beta}}
\newcommand{\bdelta}{\boldsymbol{\delta}}

\newcommand{\bgamma}{\boldsymbol{\gamma}}

\newcommand{\bSigma}{\boldsymbol{\Sigma}}

\newcommand{\var}{V}

\usepackage[plain,noend]{algorithm2e}

\def\T{{ \mathrm{\scriptscriptstyle T} }}

\def\var{\mbox{Var}}

\begin{document}

\def\spacingset#1{\renewcommand{\baselinestretch}%
{#1}\small\normalsize} \spacingset{1}


\if1\blind
{
  \title{\bf
  Data integration by combining big data and survey sample  data for finite population inference  }
  \author{Jae Kwang Kim
    \hspace{.2cm}\\
    Department of Statistics, Iowa State University\\
    and \\
    Siu-Ming Tam \\
    University of Wollongong and  Australian Bureau of Statistics}
  \maketitle
} \fi

\if0\blind
{
  \bigskip
  \bigskip
  \bigskip
  \begin{center}
    {\LARGE\bf Data integration by combining big data and survey sample  data for finite population inference  }
\end{center}
  \medskip
} \fi

\bigskip
\begin{abstract}
The statistical challenges in using big data for making valid statistical inference in the finite population  have been well documented in  literature.  These challenges are due primarily to statistical bias arising from under-coverage in the big data source to represent the population of interest and measurement errors in the variables available in the data set.  By stratifying the population into a big data stratum and a missing data stratum,   we can estimate the missing data stratum by using a fully responding probability sample, and hence the population as a whole by using a data integration  estimator.
By expressing the data integration estimator as a regression estimator, we can handle  measurement errors in the variables in big data and also in the probability sample. We also propose a fully nonparametric classification method for identifying the overlapping units and develop a bias-corrected data integration estimator under misclassification errors. Finally, we  develop a two-step regression data integration estimator to deal with measurement errors in the probability sample.  An advantage of the approach advocated in this paper is that we do not have to make unrealistic missing-at-random assumptions for the methods to work. The proposed method is applied to the real data example using 2015-16 Australian Agricultural Census data.
\end{abstract}

\noindent%
{\it Keywords:}
Calibration weighting; 
Measurement error; Non-response;  Regression estimation;   Selection bias.
\vfill

\newpage

\spacingset{1.5} 

\section{Introduction}

Suppose we are interested in estimating some finite population parameters, e.g. the finite
population mean, of a target population based on a data set.  If the data set comes from a probability sample,  parameter estimation is
straightforward,  and we can draw on the extensive literature on survey sampling over the past
century, e.g. \cite{fuller09}, \cite{sarndal92}, \cite{chambers12}. However, if the data set comes from a
non-probability sample, e.g. from a big data
source, the estimation is less straightforward, and  the theory for making inference with non-probability
samples is not fully developed. \citet{tam15} and
\citet{pfeffermann15} addressed methodological uses and challenges of big data in the production of official statistics.


The perils and pitfalls in using big data are primarily under and over coverage, and self selection.  Bias from under coverage is akin to bias from non-random samples for inference, and the bias from self-selection is akin to nonresponse bias in surveys.  These biases have been discussed extensively in the statistics literature (see for example, \cite{elliott17}, \cite{groves06}, \cite{groves08}, \cite{hand18}, \cite{kaplan14}, \cite{keiding16}, \cite{lohr17}, \cite{sax03}, and \cite{tam18}).  Specific discussion of these biases can be found in \cite{Baeza18}  for web data; \cite{brodie18} on data from smart phones and wearable devices; and \cite{olteanu19} for social media data. The weighting methods considered in \citet{valliant2011} and  \citet{elliott17}  are based on 
a missing-at-random  assumption (MAR)  of  \citet{rubin76}. 
The MAR assumption is a strong assumption and there is no way to verify this assumption from the data only.

Survey data integration, which is developed to combine information for two independent surveys from the same target population, can be used to handle the selection bias of non-probability samples by incorporating a probability sample. \citet{rivers2007}  proposed a mass imputation approach for survey integration. In \citet{rivers2007}, the nearest neighbor matching imputation is used to identify the imputed value for each element in the probability sample. \citet{zhang2012} developed a statistical theory for register-based statistics and data integration.
\citet{bethlehem16} discussed practical issues in sample matching for solving the selection bias in the non-probability sample.
While  matching-based imputation is promising and potentially useful in practice, it is still based on the missing-at-random  assumption.  \citet{chen2018} also considered a weighting  adjustment method based on parametric model assumptions on the  selection mechanism for the non-probability sample, but the MAR  assumption is still required.  \cite{rao20} provided comprehensive reviews of statistical methods of data integration  for finite population inference.

In this paper, we propose a novel method of data integration for  handling big data by incorporating survey sample data. The sampling mechanism for big data  is not necessarily MAR. That is, there can be some systematic difference between the big data sample and the survey sample even after adjusting for the auxiliary variables.  We assume that the survey variables are observed in both samples, but allow them  to be inaccurately measured in one sample. Our approach is to treat the big data sample as a finite population of incomplete (or inaccurate) observations. Furthermore, the incomplete observations in the population can be treated as auxiliary information for calibration weighting \citep{dev92,kim10b}. Thus, standard  techniques such as calibration weighting for incorporating  auxiliary information from the finite population can be used directly. To conduct calibration estimation in the survey data, we need to identify the subset of the probability sample that also belongs to the big data sample. This  is somewhat similar in spirit to dual frame estimation \citep{hartley62, skinner96}. In our application, the big data sample is subject to coverage errors, but the survey sample is not.   The proposed method is particularly useful for government statistical agencies which can effectively apply such matching.
 When the accurate matching is not possible, we propose a novel  classification method to identify the overlapping units using the matching variables observed from two data sources. Fully nonparametric propensity scores are obtained from the proposed classification procedure and  they  can be used to correct for the bias in applying the data integration estimator with inaccurate matching.


The paper is organized as follows. In Section 2, basic setup is introduced. In Section 3, the basic idea for data integration is introduced.  In Section 4, a semi-supervised classification method is introduced to identify the overlapping units with big data. In Section 5, an efficient method for data integration is introduced. In Section 6, the proposed method is extended to the case of measurement errors in the sample observation.
Two  limited simulation studies are presented in Section 7 and an application of the proposed method to an official statistics is presented in Section 8. Some concluding remarks are made in Section 9.

\section{Basic setup}

Consider a finite population $U=\{1, \cdots, N\}$ of size $N$.
From the finite population, we have two samples, denoted by $A$ and $B$,
where $A$ is  a probability sample and $B$ is a big data sample obtained by an unknown selection mechanism. From both samples, we measure the study variable $Y$. 
Initially, we assume  that $Y$ is measured without measurement error in sample $A$, but we shall relax this assumption in Section 6.
However, in  sample $B$, $Y$ is not necessarily measured accurately. Thus, instead of observing $y_i$, we observe $y_i^*$, which is a contaminated version of $y_i$, from sample B.  For simplicity, we assume that
\begin{equation}
y_i^* = \beta_0 + \beta_1 y_i + e_i,
\label{1}
\end{equation}
where $(\beta_0, \beta_1)$ is an unknown parameter and $e_i \sim (0, \sigma^2)$. Model (\ref{1}) implies that $y^*_i$ can be systematically different from $y_i$. In the special case of $(\beta_0, \beta_1)=(0,1)$,  there is no measurement bias in $y^*_i$. In addition, since the selection mechanism for the big data sample is unknown, it is subject to selection bias. Generally speaking, the selection bias of big data cannot be ignored, and adjusting for the selection bias is critical \citep{meng2018}.

To correct for the selection bias and measurement errors in the big data, we assume that we have a gold standard survey sample. Obtaining survey sample data is often expensive, but the gold standard can be used to improve the quality of the big data sample. Furthermore, optimal allocation of the resources can   make the final analysis more cost-effective.

To make sample $A$  a gold standard sample, a probability sampling design for selecting sample $A$ is employed, and $y_i$ are accurately observed from the sample. From sample $A$, we can compute $\hat{T}_a = \sum_{i \in A} d_i y_i$, a design-unbiased estimator of $T=\sum_{i=1}^N y_i$, where $d_i = \pi_i^{-1}$ is the design weight of unit $i$, and $\pi_i$ is the first-order inclusion probability of unit $i$  in sample $A$.
Table 1 presents the data structure of our setup.
We also assume that it is possible to identify  elements in sample $A$  also belonging  to sample $B$.  That is, we can create $\delta_i$ for $i \in A$, where
\begin{equation}
\delta_i = \left\{
\begin{array}{ll}
1 & \mbox{ if } i \in B \\
0 & \mbox{ otherwise. }
\end{array}
\right.
\label{2}
\end{equation}
Thus, we can observe $\delta_i$ in sample $A$ if the individual-level matching is possible. We shall relax this assumption in Section 5.

\begin{table}[!hbtp]
  \begin{center}
  \caption{Data Structure}
  \begin{tabular}{c|ccc}
  \hline
  Data    &  $Y^*$ &  $Y$ & Representative? \\
  \hline
  A    &  &  \checkmark & Yes  \\
  B   &   \checkmark &  & No \\
  \hline
\end{tabular}
  \end{center}
\end{table}

Our goal is to combine the observations in the two data sets  to find an improved estimator of $T$. By making a proper use of big data through weighting, we can obtain an improved estimator of $T$ over $\hat{T}_a$, which  completely ignores the information in the big data sample. Combining two data sources is called data integration,  and we will consider data integration as a general tool for making a proper use of big data for finite population inference.
Challenges in data integration are outlined in  \citet{lohr17} and \citet{hand18}. 
\citet{tam18} provided methods for adjusting such bias by using data integration. 
This paper extends the work  of \citet{tam18}  to non-binary variables, and also addresses situations when there are measurement errors or matching errors  in the data sets.


\section{Data integration  for handling selection bias}

We first consider the simple case of no measurement errors in $Y$, i.e.,  $y_i^* = y_i$. Now, we can conceptually define $\delta_i$ in (\ref{2}) throughout the finite population. Thus, the set of elements with $\delta_i=1$ is the big data sample. We can decompose
$$ T= \sum_{i=1}^N y_i = T_b + T_{c},  $$
where $T_b = \sum_{i =1}^N \delta_i y_i$ and $T_{c}= \sum_{i=1}^N (1-\delta_i) y_i$. Since $T_b$ can be obtained from sample $B$, we  only have to estimate $T_{c}$ from  sample $A$. Thus, we can use
\begin{equation*}
\hat{T}_{DI} = T_b + \sum_{i \in A} d_i (1-\delta_i) y_i
\end{equation*}
as a design-based estimator of $T$ obtained from two samples.  If the population size  $N$ is known,  a better estimator is
\begin{equation}
\hat{T}_{PDI} = T_b + (N-N_b) \frac{  \sum_{i \in A} d_i (1-\delta_i) y_i}{ \sum_{i \in A} d_i (1-\delta_i) }  ,
\label{3b}
\end{equation}
where $N_b= \sum_{i=1}^N \delta_i$ is the size of sample $B$.
Estimator $\hat{T}_{PDI}$ in (\ref{3b}) is essentially a post-stratified estimator with the two post-strata defined by $\delta_i=1$ and $\delta_i=0$,  respectively.

The design variance of $\hat{T}_{PDI} $ in (\ref{3b}) is 
\begin{equation*} 
 \var(  \hat{T}_{PDI} ) = (N-N_b)^2 \var\left\{ \frac{  \sum_{i \in A} d_i (1-\delta_i) y_i}{ \sum_{i \in A} d_i (1-\delta_i) } \right\} \approx \var\left\{ \sum_{i \in A} d_i (1-\delta_i) (y_i - \bar{Y}_c) \right\} ,
\end{equation*} 
where $\bar{Y}_c= \sum_{i=1}^N (1-\delta_i) y_i/ (N-N_b)$. Here, the approximate equality  follows from Taylor  linearization applied to the ratio component in (\ref{3b}).  
If the sampling design for sample  $A$ is simple random sampling of size $n$ with $n/N\approx 0$, we have
\begin{equation}
\var(  \hat{T}_{PDI} ) \approx (1-W_b) \frac{N^2}{n} S_{c}^2,
\label{4}
\end{equation}
where  $W_b=N_b/N$ and  $S_c^2 = (N-N_b)^{-1} \sum_{i=1}^N (1-\delta_i) (y_i - \bar{Y}_c)^2 $.  Thus, the variance reduction of $\hat{T}_{PDI}$ compared with  $\hat{T}_a = \sum_{i \in A} d_i y_i$ is
$$  \frac{ \var ( \hat{T}_{PDI} ) }{\var ( \hat{T}_a )} =  (1- W_b) \frac{ S_c^2}{ S^2 } . $$
If $S_c^2 \approx S^2$, the data integration estimator  is always more efficient than the design-based estimator using sample $A$ only.
In fact, from (\ref{4}), the effective sample size  using the post-stratified data integration estimator is
$$n^* = n  \frac{1}{1-W_b}   \frac{S^2}{S_c^2} . $$

Thus,  if we define $c_a$ and $c_b$ to be the per-unit cost of observing $y_i$ in sample $A$ and sample $B$, respectively,  the total cost function using post-stratified data integration estimation is
$ C_{DI} = c_a n + c_b N_b, $ while the total cost required to obtain the same efficiency of $\hat{T}_a$  is
$ C_a = c_a n^*. $
If $S_c^2 \approx S^2$, we have
$$ C_{DI} - C_a = c_b N W_b - c_a n \frac{W_b}{1-W_b} . $$
Therefore, given the same efficiency, the cost for using post-stratified data integration estimator is lower than using sample $A$ only  if
\begin{equation}
\frac{c_b}{c_a} \le \frac{n}{N } \frac{1}{1-W_b}.
\label{5}
\end{equation}
Thus,  if the under-coverage rate of $B$ is less than  $(c_a/c_b)\cdot (n/N)$ , the proposed data integration estimation  is cost-effective by  (\ref{5}).

\section{Efficient estimation}

We now discuss how to further  improve the efficiency of the data integration estimator.  One approach is to  use the idea of ratio estimation for $T$ by treating $x_i = \delta_i y_i$ as the auxiliary variable, which is observed throughout the finite population. Thus,
$$ \hat{R} = \frac{ \sum_{i=1}^N x_i}{ \sum_{i \in A}  d_i x_i } $$
can be multiplied  to direct estimator to reduce the variance, that is, to improve efficiency. The resulting ratio estimator is
\begin{equation}
\hat{T}_{RatDI}  = \hat{T}_a \hat{R} = T_b \frac{ \hat{T}_a}{ \hat{T}_b },
\label{7}
\end{equation}
where $\hat{T}_b = \sum_{i \in A} d_i \delta_i y_i$ and  $\hat{T}_a = \sum_{i \in A} d_i y_i$.  Thus, $\hat{T}_{RatDI}$ in (\ref{7}) is called the ratio data integration estimator.  Note that we can express $ \hat{T}_{RatDI}$   as
$$\hat{T}_{RatDI}  = \sum_{i \in A} d_i \left( \frac{T_b}{ \hat{T}_b} \right) y_i  = \sum_{i \in A} w_i  y_i , $$
where $w_i$ satisfies
\begin{equation}
\sum_{i \in A} w_i x_i = \sum_{i \in A} d_i \left( \frac{T_b}{ \hat{T}_b} \right) \delta_i y_i = \sum_{i=1}^N \delta_i y_i =    \sum_{i=1}^N x_i .
\label{8}
\end{equation}
Thus, equality (\ref{8}) implies that the ratio data integration estimator satisfies the calibration property of the auxiliary variable in the sense that the estimator applied to $x_i$  matches  the known population total of $x_i$.

More generally, we can  apply the calibration estimation method  to  $\bx_i = (1, \delta_i, \delta_i y_i)^{\T} $, since $\sum_{i=1}^N (1, \delta_i, \delta_i y_i ) = (N, N_b, T_b)$ is known.  Specifically, we can find $\{w_i : i\in A\}$ that minimizes an  objective  function $Q(d, w)$ subject to the calibration equation $\sum_{i \in A} w_i {\bx}_i = \sum_{i =1}^N {\bx}_i$. The  regression estimator is based on
$$ Q(d, w) = \sum_{i \in A} d_i \left( \frac{w_i}{d_i} -1 \right)^2 .
$$
The solution to the optimization problem is
\begin{equation}
w_i = d_i  \mathbf{X}_N^\T \left( \sum_{i \in A} d_i {\bx}_i {\bx}_i^\T \right)^{-1}  {\bx}_i ,
\label{9}
\end{equation}
where $\mathbf{X}_N = \sum_{i=1}^N \bx_i$.

To understand the solution in (\ref{9}), if we write ${\bx}_i =(1-\delta_i, {\bx}_{1i}^{\T})^{\T}$ with ${\bx}_{1i} = \delta_i (1, y_i)^{\T}$, the regression weight in (\ref{9}) reduces to
\begin{equation}
w_i = \left\{
\begin{array}{ll}
d_i \mathbf{X}_1^\T  \hat{\bSigma}_{xx11}^{-1} {\bx}_{1i} & \mbox{ if } \delta_i = 1 \\
d_i (N_c/\hat{N}_c) & \mbox{ if } \delta_i = 0 ,
\end{array}
\right.
\label{10}
\end{equation}
where $\mathbf{X}_1 = \sum_{i=1}^N {\bx}_{1i}$,
$  \hat{\bSigma}_{xx11} = \sum_{i \in A} d_i {\bx}_{1i} {\bx}_{1i}^\T, $ $N_c= N-N_b$  and $\hat{N}_c = \sum_{i \in A} d_i (1-\delta_i)$. The weights in (\ref{10}) satisfy
\begin{eqnarray*}
	\sum_{i \in A} w_i (\delta_i, \delta_i y_i)= (N_b, T_b), \quad
	\sum_{i \in A} w_i (1-\delta_i)  = N_c .
\end{eqnarray*}
The regression data integration estimator is then defined as
\begin{equation}
\hat{T}_{RegDI} = \sum_{i \in A} w_i y_i ,
\label{11}
\end{equation}
where $w_i$ is defined in (\ref{10}).  Inserting (\ref{10}) into (\ref{11}), we can write
\begin{equation}
\hat{T}_{RegDI} = \sum_{i =1}^N \delta_i (1, y_i)^\T \hat{\bbeta}_1 + N_c \frac{ \hat{T}_c}{ \hat{N}_c} ,
\label{11b}
\end{equation}
where $ \hat{T}_c = \sum_{i\in A}d_i(1-\delta_i)y_i$ and
$$ \hat{\bbeta}_1 =\left\{  \sum_{i \in A} d_i \delta_i (1, y_i) (1, y_i)^\T  \right\}^{-1} \sum_{i \in A} d_i \delta_i  (1, y_i)^\T y_i  =
(0,1)^\T.  $$
Therefore, the regression data integration estimator in (\ref{11b}) is algebraically equivalent to the post-stratified data integration estimator in (\ref{3b}). However, we can include other auxiliary variables observed throughout the finite population in the calibration equation; see   Remark 1 below for details.

For variance estimation, standard linearization methods or replication methods for regression estimator can be applied. For example, a linearization variance estimator for (\ref{11}) can be written as
\begin{equation}
\hat{V} ( \hat{T}_{RegDI}) = \sum_{i \in A} \sum_{j\in A} \frac{ \pi_{ij} - \pi_i \pi_j}{ \pi_{ij} } \frac{ \hat{e}_i }{ \pi_i}
\frac{ \hat{e}_j }{ \pi_j},
\label{vhat}
\end{equation}
where $\pi_{ij}$ is the joint inclusion probability of unit $i$ and $j$, $\hat{e}_i = y_i - {\bx}_i^\T \hat{\bbeta}$ and \\
$ \hat{\bbeta} =\left( \sum_{i \in A} d_i {\bx}_i {\bx}_i^\T \right)^{-1} \sum_{i \in A} d_i {\bx}_i y_i$. 

\begin{remark}
	In addition to $y_i$, if there is another variable $z_i$ observed in both samples, we can incorporate this information into calibration estimation. That is, we use ${\bx}_i = (1-\delta_i, \delta_i, \delta_i y_i, \delta_i z_i)^{\T}$ in the calibration estimation.  If $z_i$ is observed throughout the finite population, we can use ${\bx}_i = (1-\delta_i, \delta_i, \delta_i y_i,  z_i)^{\T}$.
	\end{remark}

\begin{remark}
In some cases, the big data may have duplication and lead to over-coverage problems. In this case, we can still apply the idea of calibration estimation by modifying the definition of $\delta_i$ to be the  number of times that the unit  appears in sample B. In this case, we can use
\begin{equation} 
 \sum_{i \in A} w_i ( 1, \delta_i, \delta_i y_i ) = \sum_{i \in U} (1, \delta_i, \delta_i y_i ) 
 \label{cal2} 
 \end{equation} 
as the calibration equation.
\end{remark}

\begin{remark} 
The  proposed method is also applicable  when  measurement errors exist in addition to selection bias in big data sample. That is, instead of observing $y_i$, we observe $y_i^*$, an inaccurate measurement for $y_i$, in sample $B$. In sample $A$, in addition to  observing $(y_i, \delta_i)$, we assume that  it is possible to obtain $y_i^*$ for units with $\delta_i=1$ by matching. Thus, we observe $(y_i, \delta_i, \delta_i y_i^*)$ in sample $A$.
In this case, we can still use $\delta_i y_i^*$ as a control for the calibration equation. Thus, instead of using ${\bx}_i= (1- \delta_i, \delta_i, \delta_i y_i)^{\T}$, we can use ${\bx}_i^*=(1-\delta_i, \delta_i, \delta_i y_i^*)^{\T}$ in (\ref{10}) to get the calibration weights satisfying $\sum_{i \in A} w_i (1-\delta_i) = N_c$, $\sum_{i \in A} w_i \delta_i = N_b$ and $ \sum_{i \in A} w_i \delta_i y_i^* =   \sum_{i \in B} y_i^*. $ 

\end{remark}

\section{Semi-supervised classification
}

The proposed method in Section 3 is based on the assumption that the big-data indicator function $\delta_i$ is observed for every element in sample $A$. If we have an access to the unique identifiers then it is possible to match the records accurately  and obtain $\delta_i$. In other cases, we only have matching variables such as name, zip code, and date of birth, etc. In this case, we use these matching variables to obtain the best guess of $\delta_i$, denoted by $\hat{\delta}_i$, based on the observed value of the matching variables $\mathbf{z}_i$. Obtaining $\hat{\delta}_i$ from the matching variables is a challenging classification problem. Furthermore, finding a bias-corrected estimator under misclassification error is not considered in the literature. In the context of multiple frame surveys, \cite{lohr11} developed a bias-adjustment method assuming that the misclassification probabilities are known.

In our setup, note that $\delta_i$ is observed for sample $B$, as $\delta_i=1$ if $i \in B$ by definition. We do not observe $\delta_i$  if  $i \in A$. Thus, this is a semi-supervised classification problem because the true label ($\delta_i$) for classification is available only for sample $B$. Here, we shall propose a maximum likelihood method of semi-supervised classification under the setup of data integration. Note that unlike the probabilistic record linkage, we do not have to identify the pairs of matches and non-matches as in \cite{fellegi76}. We have only to determine  whether each unit in sample $A$  belongs to the particular subpopulation $B$ or not. 

To formally describe the idea of the proposed method, recall that 
the finite population $U$ is  decomposed into two groups, $U=B \cup B^c $.  We assume that $\pi = P( \delta = 1)$ is known and given by $\pi = N_b/ N$.  We have a probability sample $A$ selected from $U$ and observe $\bz_i$ instead of observing $\delta_i $ for all $i \in A$. 
 If the densities for two groups, $p( \bz \mid \delta =1)$ and $p( \bz \mid \delta = 0)$,  are known or estimated from the samples, then we can use 
$$ P( \delta_i=1 \mid \bz_i ) = \frac{ \pi p( \bz_i \mid \delta_i  = 1) }{ (1-\pi) p( \bz_i \mid  \delta_i = 0) + \pi p( \bz_i \mid \delta_i = 1) } 
$$
to make classification for unit $i \in A$. We use $\hat{\delta}_i=1$ if we classify unit $i$ as $i \in B$.  Otherwise, we use $\hat{\delta}_i=0$. The decision rule is 
\begin{equation} 
 \hat{\delta}_i = 1 \iff \hat{P} ( \delta_i=1  \mid \bz_i ) > \frac{1}{2} , 
 \label{classification} 
 \end{equation} 
where 
$$ \hat{P}( \delta_i=1 \mid \bz_i ) = \frac{ \pi \hat{p}( \bz_i \mid \delta_i  = 1) }{ (1-\pi) \hat{p}( \bz_i \mid  \delta_i = 0) + \pi \hat{p}( \bz_i \mid \delta_i = 1) } . 
$$
 Note that $p( \bz \mid \delta =1)$ means the marginal density function of $\bz$ among big data. Estimation of $p( \bz \mid \delta =1)$ is straightforward as long as we have access to the big data. Thus, we have only to estimate  parameters in $p( \bz \mid \delta=0)$. 

To discuss parameter estimation, suppose that $\bz=(z_1, \cdots, z_K)$ and each $z_k$ can take one of $D$ values among the set $\mathcal{Z}_k= \{ z_k^{(1)} , \cdots, z_k^{(D)}  \}$ with unknown probabilities. We assume that 
    \begin{equation} 
     p( \bz \mid \delta=1) = \prod_{k=1}^K p_k ( z_k \mid \delta = 1) \label{z1}
     \end{equation} 
    where 
$    p_k (z_k \mid \delta =1) = m_{kd} \ \ \mbox{ if } z_k = z_k^{(d)} $
    and     $ \sum_{d=1}^{D} m_{kd} = 1. $ Since 
     we can observe $\bz_i$ among $\delta_i=1$, we can estimate $m_{kd}$ using 
    $$ \hat{m}_{kd} = \frac{1}{N_B} \sum_{i \in B}I  \left( z_{ik} = z_k^{(d)} \right) . $$
Now, for the model for $ p( \bz \mid \delta=0)$, we assume that 
    $$ p( \bz \mid \delta=0) = \prod_{k=1}^K p_k ( z_k \mid \delta = 0) , $$
    where 
    $p_k (z_k \mid \delta =0) = u_{kd} \ \ \mbox{ if } z_k = z_k^{(d)} $ 
    and     $ \sum_{d=1}^{D} u_{kd} = 1.$
 If we define 
    $$ \gamma_{ik}^{(d)} = 
\left\{ \begin{array}{ll} 
1 & \mbox{ if } z_{ik} = z_k^{(d)} \\
0 & \mbox{ otherwise, } 
\end{array} 
\right.
$$
    then we can express 
$ m_{kd} = P ( \gamma_{ik}^{(d)} = 1 \mid \delta_i=1)  $ and 
$ u_{kd} = P ( \gamma_{ik}^{(d)} = 1 \mid \delta_i=0) . $

 To estimate $u_{kd}$, we use the following EM algorithm: 
	\begin{enumerate} 
		\item First note that,  if $\delta_i$ were observed, the complete-sample pseudo log-likelihood would be 
		\begin{eqnarray*} 
			l_{com} (  \mathbf{u}   \mid \bdelta,  \bgamma)&=& \sum_{i \in A} d_i  \delta_i \log \left\{ \pi  \prod_{k=1}^K  m_{ik} \right\}+  \sum_{i\in A }d_i  (1- \delta_{i})  \log \left\{ (1-\pi )  \prod_{k=1}^K   u_{ik} \right\} 
		\end{eqnarray*} 
		where $$
( m_{ik}, u_{ik}) =  \sum_{d=1}^{D} \gamma_{ik}^{(d)} (m_{kd} , u_{kd}) 
.
$$
		Note that there is no need to estimate $m_{kd}$ again, because we have access to big data directly. Only $u_{kd}$ are the parameters of interest. 
		\item In the E-step, we need to evaluate the conditional expectation of $ l_{com} (  \mathbf{u}   \mid \bdelta, \bgamma)$ given the observed data. Thus, given the current parameters, we have only to compute 
		\begin{eqnarray*} 
			Q(  \mathbf{u}   \mid   \mathbf{u}^{(t)}  ) &=& E \left\{l_{com} (  \mathbf{u}   \mid \bdelta, \bgamma) \mid  \mathbf{u}^{(t)}     \right\} \\
			&=& \sum_{i\in A  } d_i \hat{p}_i^{(t)}  \log \left\{ \pi  \prod_{k=1}^K    m_{ik}  \right\} + \sum_{i\in A} d_i  (1- \hat{p}_{i}^{(t)} )  \log \left\{  (1-\pi )  \prod_{k=1}^K    u_{ik} \right\} , 
		\end{eqnarray*} 
		where 
		\begin{eqnarray} 
		\hat{p}_{i}^{(t)}  &=& E ( \delta_{i} \mid \bgamma_i ;  \hat{\mathbf{u}}^{(t)}    ) \label{gg} \\
		&=& \frac{{\pi} \prod_{k=1}^K   {m}_{ik}   }{ {\pi}\prod_{k=1}^K   {m}_{ik} +(1- {\pi})  \prod_{k=1}^K   \hat{u}_{ik}^{(t)}   } \notag 
		\end{eqnarray} 
		and 
		$ ({m}_{ik} , \hat{u}_{ik}^{(t)}  ) = \sum_{d=1}^{D} \gamma_{ik}^{(d)} ( {m}_{kd}, \hat{u}_{kd}^{(t)} ) $.  
		\item The M-step is to maximize the $Q$ over $  \mathbf{u} $ to update the parameters. The updating formula is 
		\begin{eqnarray*} 
			\hat{u}_{kd}^{( t+1)}  &=& \frac{ \sum_{i \in A}d_i (1- \hat{p}_{i}^{(t)})  \gamma_{ik}^{(d)}   }{ \sum_{i \in A} d_i (1-\hat{p}_{i}^{(t)} ) }. 
		\end{eqnarray*}  
		\item Set $t=t+1$ and go to Step 2. Continue until convergence. 
	\end{enumerate} 

Once $\hat{\delta}_i$ are computed, we may want to  use, instead of (\ref{cal2}),
\begin{equation} 
 \sum_{i \in A} w_i ( 1, \hat{\delta}_i, \hat{\delta}_i y_i ) = \sum_{i \in U} (1, \hat{\delta}_i, \hat{\delta}_i y_i ) 
 \label{cal3} 
 \end{equation} 
as the calibration equation.  The calibration estimator using (\ref{cal3}) is equivalent to 
$$\hat{T}_{PDI2} = T_{b2} + (N-N_{b2}) \frac{  \sum_{i \in A} d_i (1-\hat{\delta}_i) y_i}{ \sum_{i \in A} d_i (1-\hat{\delta}_i) }  ,
$$
where 
$(N_{b2},  T_{b2})  = \sum_{i \in U} \hat{\delta}_i(1,  y_i) $. However, unless $\hat{\delta}_i = \delta_i$, we do not observe $N_{b2}$ and $T_{b2}$ and cannot compute $\hat{T}_{PDI2}$ above. 

To overcome this difficulty, note that $\hat{p}_i$ in (\ref{gg})  is a consistent estimator of $E( \delta_i \mid \bz_i)$. Thus, as long as 
\begin{equation} 
 P( \delta=1 \mid \bz, y) = P( \delta=1 \mid \bz) 
 \label{mar} 
 \end{equation} 
holds, we can  estimate $N_{b2}$ and $T_{b2}$ consistently  by applying the standard propensity score method using $\hat{p}_i$. That is, use 
\begin{equation} 
\left( \hat{N}_{b2}, 
\hat{T}_{b2}\right)  = \sum_{i \in U} \frac{\delta_i}{ \hat{p}_i } \hat{\delta}_i(1,  y_i)    = \sum_{i \in B} \frac{\hat{\delta}_i}{ \hat{p}_i} (1, y_i ) 
\label{eqn20} 
\end{equation} 
as a propensity score  estimator of $(N_{b2},  T_{b2})  = \sum_{i \in U} \hat{\delta}_i(1,  y_i) $. Unlike \cite{chen2018}, the estimated propensity scores $\hat{p}_i$ are fully nonparametric.  
Ignoring estimation errors in $\hat{p}_i$, we have 
$$ E_{\delta} \{ ( \hat{N}_{b2},  \hat{T}_{b2} ) \} \cong E_{\delta}\left\{  \sum_{i \in U} \frac{\delta_i}{ {p}_i } \hat{\delta}_i(1,  y_i)    \right\} =  \sum_{i \in U} \hat{\delta}_i(1,  y_i)  = (N_{b2}, T_{b2} ), $$
where $E_{\delta} ( \cdot)$ denotes the expectation with respect to $\delta$ and the first equality holds because $E( \delta_i \mid \bz_i, y_i ) = p_i$.  
Thus, the resulting data integration estimator is 
\begin{equation} 
\hat{T}_{PDI2} = \hat{T}_{b2}  + (N-\hat{N}_{b2}) \frac{  \sum_{i \in A} d_i (1-\hat{\delta}_i) y_i}{ \sum_{i \in A} d_i (1-\hat{\delta}_i) }  .
\label{di4} 
\end{equation} 

The data integration estimator in (\ref{di4}) can be viewed as a calibration estimator with calibration equation 
\begin{equation} 
 \sum_{i \in A} w_i ( 1, \hat{\delta}_i, \hat{\delta}_i y_i ) = \sum_{i \in U} (1, \hat{\delta}_i\delta_i  /\hat{p}_i, \hat{\delta}_i \delta_i y_i  /\hat{p}_i ) = \left(N,  \sum_{i \in B} \hat{\delta}_i/ \hat{p}_i,  \sum_{i \in B} \hat{\delta}_i y_i/ \hat{p}_i \right),
 \label{cal4} 
 \end{equation} 
 which requires computing $\hat{p}_i$ and $\hat{\delta}_i$ for every unit in sample B. 
Condition (\ref{mar}) can be understood as  the ignorability condition of the sampling mechanism for sample $B$. This condition is not as  strong  as it might look at first. If $y$ is categorical, one can always include $y$ into $\bz$ and apply the proposed classification method. In this case, condition (\ref{mar}) is always satisfied. For continuous $y$, we may categorize $y$ first and include it into $\bz$. See the simulation study in Section 7.2 for an example.

\section{Handling measurement errors in survey data}

We now consider the case the measurement errors exist in the survey data. For example, survey data is collected annually, and the big data is available monthly. In this case, if we are interested in estimating parameters on a  monthly basis, we can treat the observed values  in the latest year from the survey data as an inaccurate measurement for $y_i$. Thus, we observe $(\delta_i,  y_i^*) $ from sample $A$ and observe $y_i$ from sample $B$. In this case, we can use the measurement error model (\ref{1}) to obtain a design-model based estimator of $T=\sum_{i=1}^N y_i$.

To estimate $T$ under measurement errors in sample $A$ and selection bias in sample $B$, we consider the following two-step approach:
\begin{description}
\item{[Step 1]} 	 Using the measurement model, estimate the parameters in $E( y_i \mid y_i^*)= m(y_i^*; \bbeta)$ and obtain mass imputation for sample $A$. That is, create $\hat{y}_i=m( y_i^*; \hat{\beta})$ for all elements in sample $A$. If the measurement error model is (\ref{1}), then we can use
	 $ \hat{y}_i = \hat{\beta}_1^{-1} ( y_i^* - \hat{\beta}_0 ) $,
	 where $(\hat{\beta}_0, \hat{\beta}_1)$ is the estimated parameter from the elements in $A \cap B$.
\item{[Step 2]}
 Apply calibration estimation  using ${\bx}_i= (1-\delta_i, \delta_i, \delta_i y_i)^{\T}$. That is, the final estimator is
 \begin{equation}
 \hat{T}_{RegDI} = \sum_{i \in A} w_i \hat{y}_i,
 \label{13}
 \end{equation}
 where $w_i$ minimizes $Q( d, w)$ subject to the calibration equation $\sum_{i \in A} w_i {\bx}_i = \sum_{i \in U} {\bx}_i$.
\end{description}

In Step 1, the bias-corrected predictor of $y_i$  is obtained from model (\ref{1}). In principle, since we observe $(y_i, y_i^*)$ among those with $\delta_i=1$ in sample $A$, we can treat this sample, $A \cap B$, as the validation sample in the calibration study. If the mechanism for $\delta_i=1$ depends on $y$ only, then the measurement error model (\ref{1}) is non-informative in the sense of \citet{Pfeffermann98}.  In this case, we can estimate model parameters in (\ref{1}) consistently by the complete-case analysis. 
That is, we can use $$ \sum_{i \in A} d_i \delta_i ( y_i^* - \beta_0 - \beta_1 y_i) (1, y_i) = (0,0) $$ as an estimating equation for $(\beta_0, \beta_1)$.

For variance estimation of  $\hat{T}_{RegDI}$ in  (\ref{13}), we can use, similarly to (\ref{vhat}), 
\begin{equation}
\hat{V} ( \hat{T}_{RegDI}) = \sum_{i \in A} \sum_{j\in A} \frac{ \pi_{ij} - \pi_i \pi_j}{ \pi_{ij} } \frac{ \hat{e}_i }{ \pi_i}
\frac{ \hat{e}_j }{ \pi_j},
\label{vhat2}
\end{equation}
where  $\hat{e}_i = \hat{y}_i - {\bx}_i^\T \hat{\mathbf{B}}$ and  
$ \hat{\mathbf{B}} =\left( \sum_{i \in A} d_i {\bx}_i {\bx}_i^\T \right)^{-1} \sum_{i \in A} d_i {\bx}_i \hat{y}_i$. Thus, we can safely ignore the effect of uncertainty of  $\hat{\bbeta}$ in $\hat{y}_i= m( y_i^*; \hat{\bbeta})$  for  variance estimation. See  Appendix A for a sketched  justification.

\section{Simulation study}
\subsection{Simulation study one}
In the first simulation, continuous $Y$ variable is considered from the following model:
\begin{equation*}
y_i  =   3+ 0.7    (x_i - 2) + e_i ,
\label{22}
\end{equation*}
where $x_i \sim N(2, 1)$, $e_i \sim N(0, 0.51)$,  and $e_i$ is independent of $x_i$. We generate a finite population of size $N=1,000,000$ from this model. 
 Also, we generate
$$ y_i^* =  2 + 0.9 ( y_i-3) + u_i $$
where $u_i \sim N(0, 0.5^2)$, and $u_i$ is independent of $y_i$.


In this simulation, we repeatedly obtain two samples, denoted by $A$ and $B$, by simple random sampling of size $n=1,000$ and by an unequal probability sampling of size $N_B=500,000$, respectively.
In selecting sample $B$, we create two strata, where stratum 1 consists of elements with $x_i\le 2$, and stratum 2 consists of those with $x_i >2$.
Within each stratum, we select $n_h$ elements by simple random sampling independently, where $n_1=300,000$ and $n_2=200,000$. Under this sampling mechanism,
the sample mean of $B$ is smaller than the population mean. We assume that the stratum information is  not available at the time of data analysis.

We consider the following three scenarios:
\begin{description}	
\item{[Scenario I]} No measurement errors in both samples. Thus, we observe $y_i$ in both samples.
\item{[Scenario II]}
	Measurement errors in sample $B$. Thus, we observe $y_i$ in sample $A$ and  $y_i^*$ in sample $B$.
\item{[Scenario III]} 	Measurement errors in sample $A$. Thus, we observe $y_i^*$ in sample $A$ and $y_i$ in sample $B$.
\end{description}
In addition, assume that we observe the matching indicator $\delta_i$ in sample $A$. If $\delta_i=1$ in sample $A$, we observe $(y_i, y_i^*)$.

We consider the following four estimators for the population mean of $Y$:
\begin{enumerate}	
\item Mean $A$. Mean of sample $A$ observations.
\item 	Mean $B$. Mean of sample $B$ observations.
\item 	Post-stratified data integration estimator of the form (\ref{3b}).
\item 	Regression data integration estimator of the form (\ref{11}).
\end{enumerate}
In Scenario II, the post-stratified data integration estimator is computed using
$$
\hat{\theta}_{PDI} =  \frac{1}{N} \left\{ \sum_{i=1}^N \delta_i y_i^*  + (N-N_b) \frac{  \sum_{i \in A} d_i (1-\delta_i) y_i}{ \sum_{i \in A} d_i (1-\delta_i) } \right\}.
$$
In Scenario III, the post-stratified data integration estimator is computed using
$$
\hat{\theta}_{PDI} = \frac{1}{N} \left\{ \sum_{i=1}^N \delta_i y_i  + (N-N_b) \frac{  \sum_{i \in A} d_i (1-\delta_i) y_i^*}{ \sum_{i \in A} d_i (1-\delta_i) } \right\},
$$
and the regression data integration estimator is computed from the two-step approach in  (\ref{13}).

\begin{table}[t]
	\caption{Results of  the four estimators for simulation study one  based on a Monte Carlo sample of size $1,000$   }
	\begin{center}
		\begin{tabular}{c|c|rcc}
\hline
			 Scenario & Estimator & Bias & SE &  RMSE   \\
		\hline
			\multirow{4}{*}{I}& Mean $A$ & 0.00 & 0.031 & 0.031 \\
			 & Mean  $B$ & -0.11 & 0.001 & 0.113\\
			& PDI &  0.00 & 0.022 & 0.022\\
			& RegDI & 0.00 &  0.022 & 0.022 \\
		\hline
			\multirow{4}{*}{II}& Mean $A$ & 0.00 & 0.031 & 0.031\\
		 & Mean $B$ & -1.10 & 0.001 & 1.101\\
			& PDI & -0.49      & 0.022 & 0.495   \\
			& RegDI & 0.00     & 0.024 & 0.024  \\
		\hline
			\multirow{4}{*}{III}& Mean $A$ & -1.00 & 0.033 &1.001  \\
		  & Mean $B$ & -0.11 &  0.001 & 0.113 \\
			& PDI & -0.51 & 0.023 & 0.507 \\
			& RegDI & 0.00 & 0.028 & 0.028  \\
		\hline
		\end{tabular}
	\end{center}
	\label{tab:m1}

SE, standard error; RMSE, root mean squared error; PDI, Post-stratified data integration estimator; RegDI,  regression data integration estimator.
	
\end{table}

Table 2 presents the result of the  simulation study based on $1\,000$ Monte Carlo samples.
From  Table 2,  mean $A$ estimator is unbiased except for Scenario III, where systematic measurement errors exist in sample $A$.  Mean $B$ estimator is always biased due to the selection bias in sample $B$. The bias is the largest in absolute values for Scenario II, where measurement errors exist in addition to the  selection bias. Variance of mean $B$ estimator is the smallest because of the large sample size of sample $B$ ($N_B=500,000$). The post-stratified data integration estimator is unbiased in Scenario I, which is consistent with our theory in Section  3.
The variance of the post-stratified estimator  is about half of the variance of the mean $A$ estimator because $N_B/N=0.5$. If the rate $W_B=N_B/N$ is larger, then the variance estimator  post-stratified estimator will be smaller as equation (4) may suggest. 
However, in Scenario II, the post-stratified data integration  estimator is biased because $T_b=\sum_{i=1}^N \delta_i y_i$ is estimated  without correcting for the measurement errors. In Scenario III, it is biased because $T_c=\sum_{i=1}^N (1-\delta_i) y_i$ is estimated from sample $A$ without correcting for the measurement errors.           The regression data integration estimator is unbiased for all scenarios. It is the same as the post-stratified data integration estimator under Scenario I, as   discussed in (\ref{11b}).

In addition, we also compute variance estimators of the regression data integration estimator using formula (\ref{vhat2}). For example, in Scenario 2, we use 
    $$ \hat{e}_i = \left\{ 
    \begin{array}{ll} 
    y_i- (\hat{b}_0 + \hat{b}_1 y_i^*)  & \mbox{ if } \delta_i = 1 \\
    y_i - \bar{y}_c & \mbox{ if } \delta_i = 0 , 
    \end{array} 
    \right.
    $$
    where $(\hat{b}_0, \hat{b}_1)$ is the solution to 
$     \sum_{i \in A} d_i \delta_i ( y_i - b_0 - b_1 y_i^* ) (1, y_i^*) = (0,0) $. Based on $1,000$ Monte Carlo samples, we compute the relative biases of the variance estimators. The relative biases are -0.0037, 0.028, and 0.019 for Scenarios 1, 2, and 3, respectively. Thus, we  conclude that the proposed variance estimators are nearly unbiased.  

\subsection{Simulation study two}

In the second simulation study, we study the performance of the data integration estimator using matching variables. 
 In the simulation study, we first generate a finite population with $(z_{i1}, z_{i2}, \delta_i, y_i)$ as follows. First generate 
$$ z_{1i} \sim \mbox{Unif} \{1, \cdots, 20 \} $$
independently. Given $z_{1i}$, we generate $\delta_i$ from Bernoulli distribution with the probability 
$$ P ( \delta_i=1 \mid z_{1i}  )  
= \left\{ \begin{array}{ll}  
c  \mbox{ if } z_{i1} \le 10 \\
2 c \mbox{ if } z_{i1} > 10 
\end{array} 
\right. 
$$
where $c$ is chosen such that the sum of the probabilities over the finite population  is equal to $N_B$. We set $N=10,000$ and $N_B=5,000$ in this simulation. 
 We also generate 
 $$ y_i = \left\{ 
 \begin{array}{ll} 
 4+ 0.5 (z_{i2}+e_i) & \mbox{ if } z_{1i} \le 10   \\
  6+ 0.3 (z_{i2}+e_i) & \mbox{ if } z_{1i} > 10   \\
 \end{array} 
 \right. 
 $$
 where
$ z_{2i}  \sim \mbox{Unif} \{1, \cdots, 10\} $, $e_i \sim \mbox{Unif}(0,1)$ and $z_{2i}$ and $e_i$ are mutually independent. Thus, we can treat $z_{2i}$ as a categorization of  continuous variable $y_i$. 

From the finite population, we select sample $A$ by simple random sampling of size $n_A$. 
Two values of $n_A=|A|$ are considered: $n_A=1,000$ versus $n_A=2,000$. From sample $A$, we observe $(z_{i1}, z_{i2}, y_i)$ but not $\delta_i$. Thus, we apply the semi-supervised classification method using $(z_{1i}, z_{i2})$ as the matching variable. Note that as $z_{i2}$ is included in the matching to satisfy the ignorability condition  (\ref{mar}) approximately.

From 
each sample, we consider five  estimators of $\bar{Y}_N =N^{-1} \sum_{i=1}^N y_i$. 
\begin{enumerate}	
\item Mean $A$. Mean of sample $A$ observations.
\item 	Mean $B$. Mean of sample $B$ observations.
\item 	Naive data integration (DI)  estimator: Treat $\hat{\delta}_i$  as if accurate and apply the data integration estimator using $\hat{\delta}_i$ to get 
$$ \hat{T}_{PDI} = T_B + \left( N- N_b \right) \frac{ \sum_{i \in A} d_i (1- \hat{\delta}_i) y_i}{ \sum_{i \in A }  d_i (1- \hat{\delta}_i)} . 
$$
\item The proposed data integration estimator: 
$$\hat{T}_{PDI2} = \hat{T}_{b2}  + (N-\hat{N}_{b2}) \frac{  \sum_{i \in A} d_i (1-\hat{\delta}_i) y_i}{ \sum_{i \in A} d_i (1-\hat{\delta}_i) }  ,
$$
where $\hat{T}_{b2}$ and $\hat{N}_{b2}$ are defined in (\ref{eqn20}).

\item The original data integration estimator  using the true indicator function $\delta_i$. This estimator is computed as a benchmark for comparison. 
\end{enumerate}

\begin{table}[ht]
\caption{Results of the five estimators for simulation study two based on a Monte Carlo sample of size 1,000}
\begin{center} 
\begin{tabular}{clccr}
  \hline
$n_A$ & Estimator & Bias  &SE & RMSE   \\
\hline 
 & Mean $A$ & 0.00 & 0.037  & 0.037  \\
&  Mean B & -0.14 & 0.011 & 0.135  \\
1,000 & Naive DI & 0.12 & 0.036  & 0.130\\
& Proposed DI & 0.00 & 0.035  & 0.035 \\
& Original DI & 0.00 & 0.024 & 0.024  \\
\hline 
 & Mean $A$  & 0.00 & 0.025 & 0.025 \\ 
  & Mean $B$  & -0.14 & 0.011 & 0.135 \\
 2,000 & Naive DI  & 0.14 & 0.015 & 0.136 \\
  & Proposed DI  & 0.00 & 0.023  & 0.023 \\ 
 & Original DI   & 0.00 & 0.016 & 0.016  \\
   \hline 
\end{tabular}
\label{table4}
\end{center} 
SE, standard error; RMSE, root mean squared error. 

\end{table}

Table \ref{table4} presents the performance of the five estimators. Mean B is seriously biased due to its selection bias. Naive DI estimator is also biased seriously due to the misclassification errors in $\hat{\delta}_i$. The proposed DI estimator is unbiased and is more efficient than the sample mean (Mean A) of the sample A, although the efficiency gain is not as significant as in the original DI estimator. The efficiency gain will increase with $W_b=N_b/N$.



\section{An Application in Official Statistics}

We now consider an application of the proposed method to a real data problem  using  2015-16 Australian Agricultural Census as the big data, which has 85\% response rate. In addition, we use the 2014-15 Rural Environment and Agricultural Commodities Survey (REACS) as the probability sample (sample A)  for calibration. Our interest is to combine the Agricultural Census data with the REACS data to estimate  the total area of holdings (AOH), the total number of dairy cattle (DAIRY), the number of beef cattle (BEEF), and the number of tonnes of wheat for grain or seed produced (WHEET) for 2015-16. Thus, we observe $y_i$ from the Agricultural Census data and observe $y_i^*$ from REACS.

To apply the proposed method, define $\delta_i=1$ if unit $i$ participated at the Census and $\delta_i=0$ otherwise. Thus, in REACS sample, we observe $y_i$ in addition to $y_i^*$ for $\delta_i=1$. Using the matched sample in sample A, we can fit a measurement error model
$$ {y}_i^* = \beta_0 + \beta_1 {y}_i + u_i $$
and obtain $\hat{y}_i =   \hat{\beta}_1^{-1} ( y_i^* - \hat{\beta}_0 ) $ for all $i \in A$. Here, $y_i$ is the true value of the study variable 
from 2015-2016 Census 
and $y_i^*$ is its proxy value obtained from  2014-2015 REAC survey data.

For each parameter, we compute the following three estimators:
\begin{enumerate}
\item Survey estimate (from REACS sample): $\hat{\theta}_{HT} = \sum_{i \in A} w_i \hat{y}_i $
\item Big data estimate (from Census):  $\hat{\theta}_B = \sum_{i \in B} y_i $
\item The two-step Data integration estimate using calibration weighting:
$$ \hat{\theta}_{DI} = \sum_{i \in A} w_{i, cal} \hat{y}_i $$
where $w_{i, cal}$ satisfies
$ \sum_{i \in A}w_{i, cal} ( 1- \delta_i  , \delta_i, \delta_i x_i ) = \sum_{i \in U} ( 1- \delta_i  , \delta_i, \delta_i x_i ) $
and $x_i$ includes major study variables.
\end{enumerate}
The estimates are compared with the official numbers of the Australian Bureau of Statistics (ABS), which is obtained by applying imputation for item nonresponse in the Census.

$$ < \mbox{Figure 1 around here} > $$

$$ < \mbox{Figure 2 around here} > $$

Figure 1 and Figure 2 present the estimation results for AOH and DAIRY, respectively,  by eight states in Australia.
We do not report the results for other commodities to save space. The confidence intervals are constructed using the asymptotic normality with 90\% nominal coverage rates.  The results in Figure 1 and Figure 2 can be summarized as follows: (1) The Big data estimates show serious negative biases due to the undercoverage of the big data (nonresponse in the Census), (2) The proposed data integration estimator shows narrower confidence intervals than the survey estimate,  (3) The effect of calibration weighting is reduced because of the measurement errors in sample A observations.
 Overall, the confidence intervals obtained from the proposed data integration estimators cover the official ABS estimates.

\section{Discussion}

The proposed data integration methods  feature an independent probability sample for estimating the missing data stratum of the finite population, which can correct for the under-coverage bias of the big  data sample.  By treating big data as an incomplete sampling frame for the finite population, we can apply the calibration weighting method.
 In addition,  these methods are extended in this paper to handle measurement errors in either the Big Data source or the probability sample source.   Also, a fully nonparametric approach to propensity score estimation for big data sample participation is developed using a new  semi-supervised classification method. 

In practice, our methods are useful provided the following conditions apply:
\begin{enumerate}
\item Existence of a probability sample $A$ which also measures $y$ or provides a proxy measure $y^*$. Whilst the coincidental existence of such a sample is rare, where one, e.g. a national statistical offices, determines the benefits in using big data for inference outweighs the costs, one can design, develop and implement such a random sample to collect the measure of interest.  Where this occurs, the population count of the sample units, $N$,  is by definition known.
        \item The calibration method is useful only if the coverage of $B$ is substantial, which is not an unreasonable assumption if $B$ is a big data set.  Also, when $B$ is big, it can be assumed that $A \cap B$ is not empty for measurement error  adjustment, where warranted;
\end{enumerate}

The nonparametric propensity scores  obtained from the semi-supervised classification method can be used to correct for the  coverage bias in big data sample. How to make valid statistical inference, including variance estimation, under the nonparametric propensity score adjustment is not pursued here and will be covered elsewhere.  Extensions to  small domain estimation \citep{rao2015} and analytic inferences using big data  will also be  future research topics.

\section*{Acknowledgements}

The authors are grateful  to two anonymous referees and the co-editor for the very constructive comments. 
The research of the first author was partially supported  by a grant from US National Science Foundation (MMS-1733572). 
\newpage

\appendix
 \renewcommand{\theequation}{A.\arabic{equation}}
 \setcounter{equation}{0}

\section*{Appendix}
\subsection*{A. Justification for (\ref{vhat2}) }

Let $\theta=N^{-1} T$, the finite population mean of $Y$,  be the parameter of interest.
We first consider variance estimation of the mass imputation estimator of the form
$$ \hat{\theta}_{DI} = \frac{1}{N} \sum_{i \in A} d_i \hat{y}_i,  $$
where $\hat{y}_i$ is a predictor of $y_i$ using $y_i^*$. We use $\hat{y}_i = \hat{m}^{-1} ( y_i^*) $ where $\hat{m} ( y_i )= m( y_i ; \hat{\bbeta}) = E( y_i^* \mid y_i ; \hat{\bbeta}) $. The estimating equation for $\hat{\bbeta}$ can be written as 
\begin{equation}
\hat{U}_{\beta} ( \bbeta) = N^{-1} \sum_{i \in A} d_i \delta_i \{ y_i^* -   m( y_i ; \bbeta) \} \bh( y_i ; \bbeta) = 0 
\label{a0}
\end{equation}
for some $\bh( y; \bbeta)$ such that $\hat{U}_{\beta} ( \bbeta) $ is linearly independent. 
Writing
$\hat{\theta}_{DI} = \hat{\theta}_{DI} ( \hat{\bbeta}) $,  we can use Taylor linearization   to estimate the variance of $\hat{\theta}_{DI}$.
Using the standard argument \citep{kimrao09},  we can obtain
\begin{equation}
\hat{\theta}_{DI} =  \hat{\theta}_{DI} ( {\bbeta}_N ) - E\left\{ \frac{\partial}{ \partial \bbeta^\T}  \hat{\theta}_{DI} ( {\bbeta}_N )  \right\}
\left[ E\left\{ \frac{\partial}{ \partial \bbeta^\T}  \hat{U}_\beta ( {\bbeta}_N )  \right\} \right]^{-1} \hat{U}_\beta( \bbeta_N ) + o_p (n^{-1/2}) ,
\label{a1}
\end{equation}
where   $\bbeta_N$ is the probability limit of $\hat{\bbeta}$. 

After some algebra,  we can express (\ref{a1}) as
\begin{equation}
\hat{\theta}_{DI} = \frac{1}{N} \sum_{i \in A} d_i \left\{ q_i  + \delta_i \left( y_i^* -m (y_i; \bbeta)   \right)  \kappa^\T \bh_i    \right\} + o_p( n^{-1/2} ) .
\label{a3}
\end{equation}
where $q_i = q_i (\bbeta_N)  $ is the solution to $ y_i^*  = m(q_i ; \bbeta_N)$, $\bh_i = \bh (y_i; \bbeta_N) $ and $\kappa$ satisfies 
$$
\sum_{i=1}^N \delta_i \dot{m}_i \bh_i^\T \kappa = \sum_{i=1}^N  \dot{q}_i 
$$
with  $\dot{m}_i = \partial m(y_i; \bbeta) / \partial \bbeta $ and $\dot{q}_i = \partial q_i( \bbeta)/ \partial \bbeta$. 
Using (\ref{a3}),  we can express
\begin{equation}
\hat{\theta}_{DI} - \theta =( \bar{q}_N - \theta ) + (\bar{u}_{HT} - \bar{u}_N)  + o_p (n^{-1/2}),
\label{a4}
\end{equation}
where $\bar{q}_N = N^{-1} \sum_{i=1}^N q_i$,  $u_i =  q_i+ \delta_i \left\{  y_i^* -  m(y_i; \bbeta_N )  \right\}  (   \kappa^\T \bh_i  ) $, $\bar{u}_{HT} = N^{-1} \sum_{i \in A} d_i u_i$ and $\bar{u}_N = N^{-1} \sum_{i=1}^N u_i $.

From (\ref{a4}), we can obtain
\begin{equation}
\var \left( \hat{\theta}_{DI} - \theta  \right)
= \var ( \bar{q}_N - \theta ) + \var(\bar{u}_{HT} - \bar{u}_N) = V_1 + V_2 .
\label{a5}
\end{equation}
The first term is of order $O(N^{-1})$, and the second term is $O(n^{-1})$. The first term is negligible if $n/N=o(1)$.
To estimate the second term of (\ref{a5}), we can use
\begin{equation}
\hat{V}_2 = \frac{1}{N^2} \sum_{i \in A} \sum_{j \in A} \frac{ \pi_{ij} - \pi_i \pi_j}{ \pi_{ij} }
\frac{ \hat{u}_i }{\pi_i } \frac{ \hat{u}_j }{\pi_j} ,
\label{var}
\end{equation}
where
$$ \hat{u}_i = \hat{q}_i  + \delta_i \{  y_i^* - m ( y_i; \hat{\bbeta})  \} ( \hat{\kappa}^\T \hat{\bh} _i )  $$
and
$$ \hat{\kappa} = \left\{ \sum_{i \in A} d_i \delta_i \dot{m}_i \bh_i^\T\right\}^{-1}  \sum_{i \in A} d_i   \dot{q}_i . 
$$

Next, we consider variance estimation for the calibration estimator 
$$\hat{\theta}_{RegDI}=\frac{1}{N}  \sum_{i \in A} w_i \hat{y}_i , $$
where $w_i$ are the calibration weights satisfying the calibration equation $\sum_{i \in A} w_i {\bx}_i = \sum_{i =1}^N {\bx}_i$.    In this case, the linearization in (\ref{a3}) changes to 
\begin{equation}
\hat{\theta}_{RegDI} = \frac{1}{N} \sum_{i \in A} d_i \left\{ e_i  + \delta_i \left( y_i^* -m (y_i; \bbeta_N )   \right)  \kappa_2^\T \bh_i    \right\} + o_p( n^{-1/2} ) , 
\label{a6}
\end{equation}
where $e_i = q_i - {\bx}_i^{\T} \mathbf{B}$, $ \mathbf{B}  = \left( \sum_{i =1}^N {\bx}_i {\bx}_i^\T \right)^{-1}  \sum_{i =1}^N \bx_i  q_i  $ and 
$\kappa_2$ satisfies 
$$
\sum_{i=1}^N \delta_i \dot{m}_i \bh_i^\T \kappa_2 = \sum_{i=1}^N e_i . 
$$
Since $\bx_i$ includes an intercept term, we have $\sum_{i=1}^N e_i =0$, which implies $\kappa_2=0$. 
 Therefore, for variance estimation of $\hat{\theta}_{RegDI}$, we can use (\ref{vhat2}), where
$ \hat{e}_i =\hat{q}_i - {\bx}_i^\T \hat{\mathbf{B}}  $
and $ \hat{\mathbf{B}} = \left( \sum_{i  \in A} d_i {\bx}_i {\bx}_i^\T \right)^{-1}  \sum_{i \in A} d_i {\bx}_i \hat{q}_i . $

\bibliographystyle{chicago}

\bibliography{bigdata2}

\newpage
\begin{figure}[!h]
	\centering
		\caption{Three estimates for AOH for 2015-16 by States }\label{AOH}

	\includegraphics[scale=1.0]{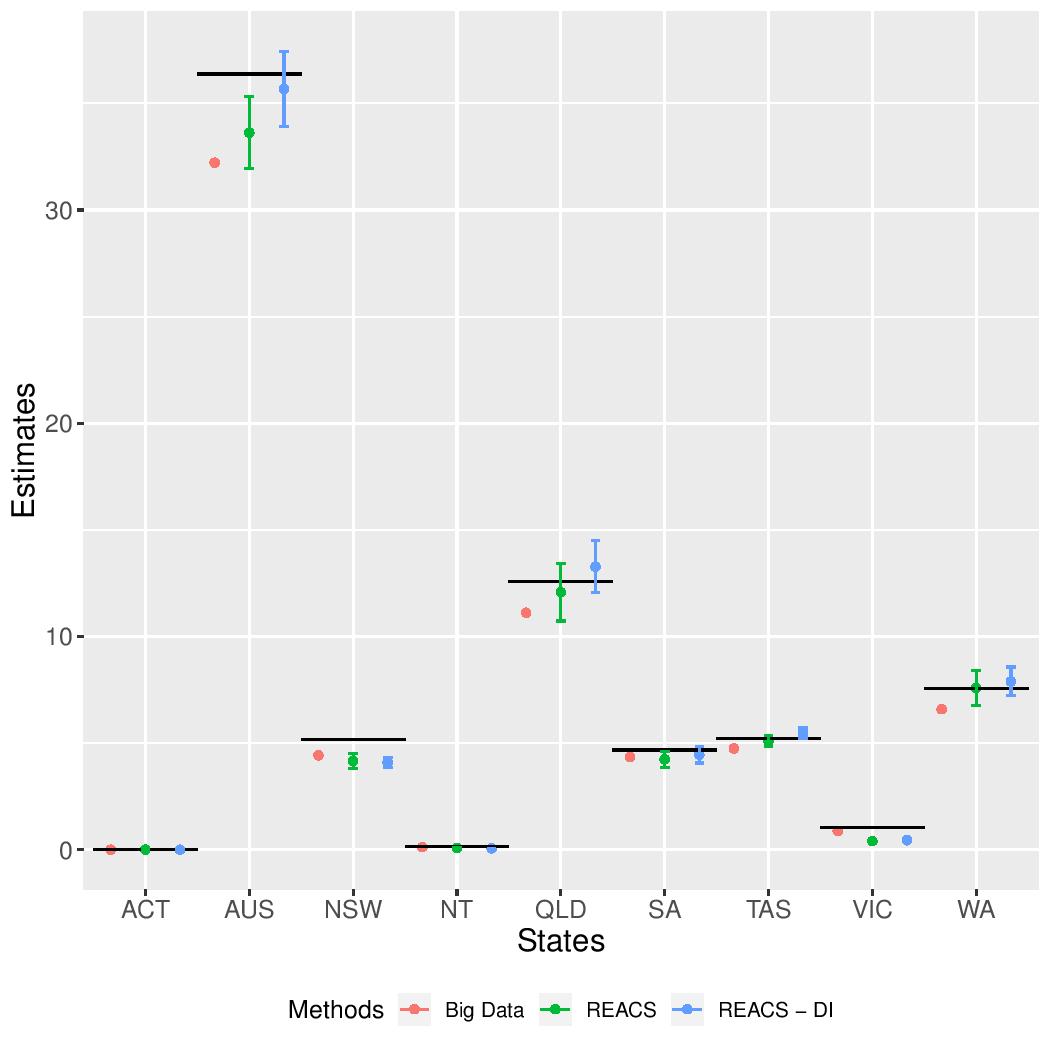}

\end{figure}

\begin{figure}[!h]
	\centering
		\caption{Three estimates for DIARY for 2015-16 by States }\label{DIARY}

	\includegraphics[scale=1.0
 ]{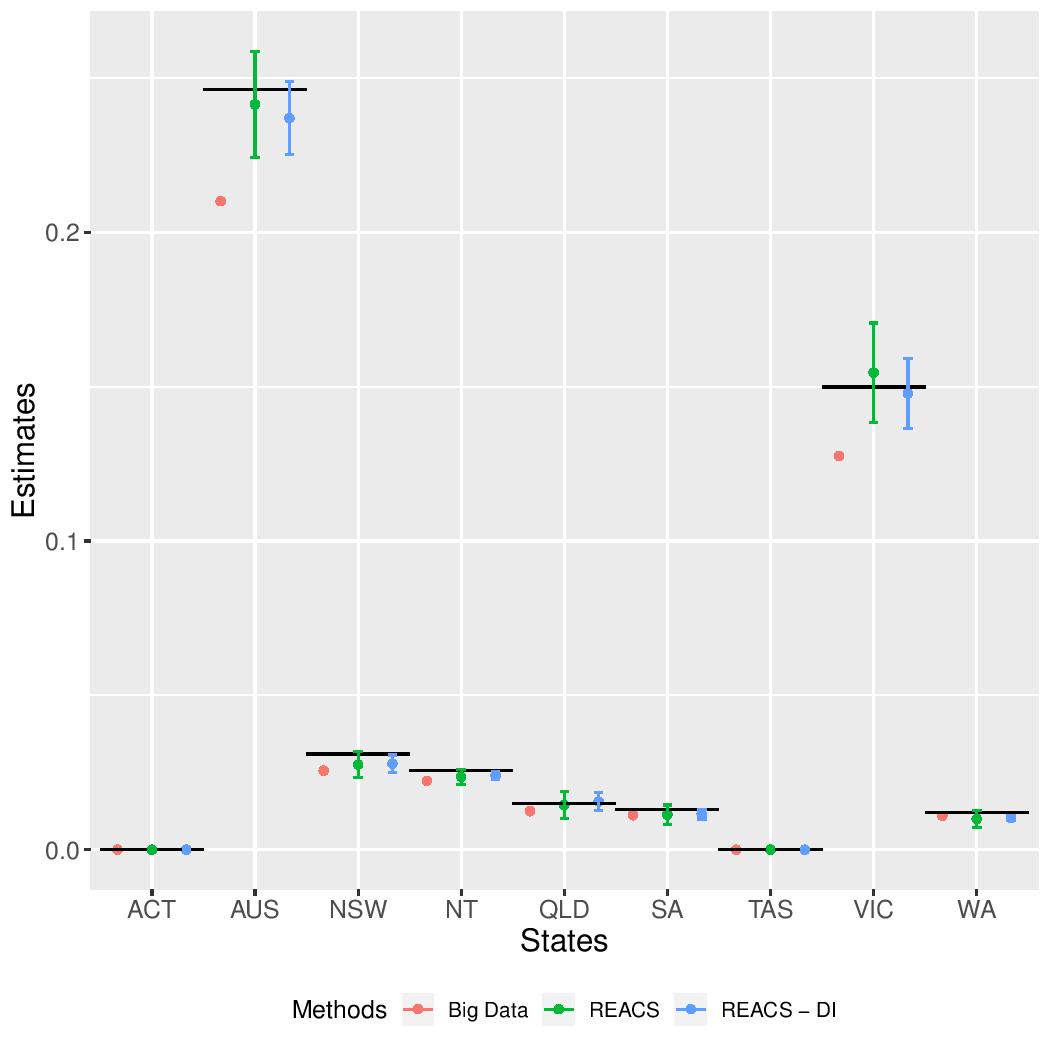}

\end{figure}

\end{document}